\documentclass[english,12pt]{article}
\usepackage[T1]{fontenc}
\usepackage[latin1]{inputenc}
\usepackage{babel}
\usepackage{cite}

\begin{document}

\title{\textbf{Energetics of Protein Thermodynamic Cooperativity: Contributions
of Local and Nonlocal Interactions}}
\author{Michael Knott, H\"useyin Kaya and Hue Sun Chan\footnote
{E-mail: chan@arrhenius.med.toronto.edu; Tel: +1~416~978~2697; Fax:
+1~416~978~8548; Mailing address: Department of Biochemistry,
University of Toronto, Medical Sciences Building, 1 King's College
Circle, Toronto, Ontario M5S 1A8, Canada.} \\
\normalsize{Protein Engineering Network of Centres of Excellence (PENCE),}\\
\normalsize{Department of Biochemistry, and} \\
\normalsize{Department of Medical Genetics \& Microbiology,} \\
\normalsize{Faculty of Medicine, University of Toronto,} \\
\normalsize{Toronto, Ontario M5S 1A8, Canada}}

\maketitle

\noindent
\textbf{Keywords:} calorimetry / G\=o models / two-state cooperativity/\\ 
single-domain proteins / radius of gyration

\section*{Abstract}

The respective roles of local and nonlocal interactions in the
thermodynamic cooperativity of proteins are investigated using
continuum (off-lattice) native-centric G\=o-like models with
a coarse-grained C$_\alpha$ chain representation. We study a
series of models in which the (local) bond- and torsion-angle terms
have different strengths relative to the (nonlocal)
pairwise contact energy terms. Conformational distributions in
these models are sampled by Langevin dynamics. Thermodynamic
cooperativity is characterized by the experimental criteria
requiring the van't Hoff to calorimetric enthalpy ratio
$\Delta H_{\rm vH}/\Delta H_{\rm cal}\approx 1$ (the calorimetric
criterion), as well as a two-state-like variation of the average
radius of gyration upon denaturation. We find that both local
and nonlocal interactions are critical for thermodynamic cooperativity.
Chain models with either much weakened local conformational propensities 
or much weakened favorable nonlocal interactions are significantly less 
cooperative than chain models with both strong local propensities and 
strong favorable nonlocal interactions. These findings are compared with 
results from a recently proposed lattice model with a local-nonlocal 
coupling mechanism; their relationship with experimental measurements of 
protein cooperativity and chain compactness is discussed.

\pagebreak
 
\section{Introduction}

How a globular protein can fold reliably into a particular three dimensional 
conformation \emph{in vitro}, without the participation of molecular 
chaperones, is a central puzzle in biophysics. If we wish not only to 
predict the folded state of a protein, but also to understand the folding 
phenomenon in terms of physical processes, we need to use physics-based 
methods: we run computer simulations of self-contained polymer models
\cite{Cha02}
that attempt to mimic the behavior of real protein molecules. A complete
quantum mechanical simulation, which would include the solvent molecules
in addition to all the atoms in the protein molecule, is not yet possible.
But in attempting to design simplified models, we face the problem
of how to simplify: which characteristics are essential and which
can be neglected? What effective energy functions does this imply
for the simplified system? 

Part of this general question is addressed in this article. Folding 
experiments on small globular proteins have long shown evidence of 
thermodynamic and kinetic cooperativity \cite{Jac91,Bak00}, which indicates a
phenomenon similar to a first order phase transition between native and 
denatured states. As our group has argued recently
\cite{Cha00,Kay00a,Kay00b,Kay02,Kay03a}, this 
observation 
can be exploited to constrain the set of possible simplified models and 
interaction schemes: for a particular simplified model to be a quantitatively
accurate representation of protein thermodynamics and kinetics, it is
essential that, when appropriately applied to a small globular protein,
it can produce the experimentally observed generic cooperative behavior.

Such constraints turn out to be rather stringent. It is nontrivial
to construct model interaction schemes that can produce proteinlike
cooperativities \cite{Cha00,Kay00a,Kay00b,Kay02,Kay03a}. 
A case in point is a class of common G\=o-like \cite{Tak75} 
models \cite{Kay03a,Cle00}. Their potential functions are native-centric, in
that they are 
explicitly biased to favor a given native structure. G\=o-like modeling
of proteins has provided important physical insights 
\cite{Cle00,Mic99,Zho99,Por01}. These include
an increasing number of elegant elucidations of functional protein dynamics 
under native conditions \cite{Hal97,Kes02,Isi02,Mic02,Jac01}. 
As for global folding and unfolding of 
proteins (in contrast to their near-native dynamics), a detailed discussion 
of the merits and limitations of G\=o-like approaches can be found in 
Ref.~\cite{Kay03a}. Notably, common G\=o-like models do not appear capable of
producing
simple two-state folding/unfolding kinetics. Instead, their chevron plots 
exhibit severe rollovers, which are typical of the class of folding
kinetics that is 
customarily referred to as non-two-state \cite{Kay03a,Kay03b}. 
Nonetheless, common
three-dimensional G\=o-like protein models seem sufficient to produce 
apparent two-state thermodynamic behavior \cite{Kay00a,Kay03a}, although their 
two-dimensional counterparts fail to do so \cite{Cha00}.

In the present investigation, we limit our scope to thermodynamic
cooperativity. Specifically, we aim to explore how protein 
thermodynamic two-state-like behavior is affected by the relative
strengths 
of local interactions (between residues close together along the chain 
sequence) and nonlocal interactions (between 
residues far apart along the chain sequence). The respective roles
of local and nonlocal interactions are an issue of long standing interest
in the study of protein energetics \cite{Go_78,Dil90,Abk95,Cha98,Bal99,Por01-2},
and the effect of analogous interactions on the phase diagram of lattice
polymers has also been investigated \cite{Deg75,Don96,Bas97,Doy98}. 
Here the issue is addressed
by varying the potential function in a series of coarse-grained G\=o-like 
models, which represent the protein as a string of C$_\alpha$ positions
in continuum space and which are simulated using Langevin dynamics. In
view
of the limitations of common G\=o models \cite{Kay03a}, the present study
should
be viewed as a first step in tackling the issue of local vs. nonlocal
interactions in cooperative continuum protein models. To assess
the robustness of our conclusions, results from the continuum Langevin
models 
are also compared with results from lattice model simulations.

We begin in section 2 by providing details of the models. An outline of the 
thermodynamics involved in interpreting the simulations is given
in section 3. Our findings are presented in section 4, and we conclude 
in section 5 with a discussion of the implications of our results.
\\

\section{Models and simulation details}

\subsection{Continuum models}

For the present continuum G\=o-like models we use a representation,
introduced by Clementi \emph{et al.} \cite{Cle00}, of the 64-residue truncated
form of chymotrypsin inhibitor 2 (CI2). The native contact set 
corresponds to NCS2 in Ref.~\cite{Kay03a}.

We use an energy function that is similar to one used previously 
\cite{Kay03a,Cle00,Kog01}.
The potential energy function \( V \), from which the conformational force
is derived, is given by 
\begin{equation}
\label{fda}
V=V_{\rm stretching}+V_{\rm bending}+V_{\rm torsion}
+V_{\rm native}+V_{\rm nonnative},
\end{equation}
where\begin{equation}
\label{fdb}
V_{\rm stretching}=\sum _{i=1}^{N-1}k_{l}\left( l^{i}-l^{i}_{0}\right) ^{2}
\end{equation}
contains a summation over the virtual bonds between pairs of residues,
\begin{equation}
\label{fdc}
V_{\rm bending}=\sum _{i=1}^{N-2}\varepsilon _{\theta }(\theta ^{i}-\theta
^{i}_{0})^{2}
\end{equation}
involves a summation over the virtual bond angles between triplets of residues,
and
\begin{equation}
\label{fdd}
V_{\rm torsion}=\sum _{i=1}^{N-3}\left\{ \varepsilon _{\phi }^{(1)}\left[ 1-\cos
\left( \phi ^{i}-\phi ^{i}_{0}\right) \right] +\varepsilon _{\phi }^{(3)}\left[ 1-\cos 3\left( \phi ^{i}-\phi ^{i}_{0}\right) \right] \right\} 
\end{equation}
represents the virtual torsional potential between quadruplets of residues.
The latter contains a term with a single minimum as well as the traditional
three-minimum term \cite{Bra67}. 
[We note that there is an apparent typographical error in the corresponding
$V_{\rm torsion}$ in \cite{Cle00}, which effectively lists these terms
as $1+\cos\left( \phi ^{i}-\phi ^{i}_{0}\right)$ and 
$1+\cos 3\left( \phi ^{i}-\phi ^{i}_{0}\right)$. But such terms would
fold the chain into the mirror image of the PDB structure.]
\( V_{\rm stretching} \), \( V_{\rm bending} \)
and \( V_{\rm torsion} \) together account for the local interactions
(between residues that are separated by no more than three places
along the chain), which include local conformational propensities for
the native structure. The local interactions are expressed in this way
because it biases the local geometry of the chain. The fourth term,
\begin{equation}
\label{fde}
V_{\rm native}=\sum _{\left| i-j\right| \geq 4}\varepsilon
_{\rm native}\left[5\left(
\frac{r_{0}^{ij}}{r^{ij}}\right) ^{12}-6\left( \frac{r_{0}^{ij}}{r^{ij}}\right)
^{10}\right], 
\end{equation}
sums over the pairwise interactions between residues that are regarded
as being in contact in the native structure; this accounts for the
nonlocal interactions (between residues that are separated by four
or more places along the chain). Finally,
\begin{equation}
\label{fdf}
V_{\rm nonnative}=\sum _{\left| i-j\right| \geq q}
\varepsilon \left( \frac{r_{\rm rep}}{r^{ij}}\right) ^{12}
\end{equation}
contains repulsive pairwise interactions between other pairs of residues,
in order to ensure the self-avoidance of the chain. 

The chain contains \( N \) residues. \( l^{i} \) is the length of
virtual bond \( i \), \( \theta ^{i} \) is a bond angle, \( \phi ^{i} \)
is a dihedral angle and \( r^{ij} \) is the distance between two
residues \( i \) and \( j \). The corresponding values in the native
structure are \( l^{i}_{0} \), \( \theta ^{i}_{0} \), \( \phi ^{i}_{0} \)
and \( r^{ij}_{0} \). The range \( r_{\rm rep} \) of the repulsive potential
between pairs of residues that are not bonded and do not interact
via a native contact interaction is set to 4~\AA. Length is expressed
in units of \AA, and energy in units of \( \varepsilon  \),
the energy parameter of the nonnative repulsive interaction, so that
\( \varepsilon  \) itself is unity. \( k_{l} \), \( \varepsilon _{\theta } \),
\( \varepsilon _{\phi }^{(1)} \), \( \varepsilon _{\phi }^{(3)} \)
and \( \varepsilon _{\rm native} \) are also parameters of the potential energy 
function. \( k_{l} \) is fixed at \( 100\varepsilon /\textrm{\AA} \),
but the other parameters can be varied; 
\( \varepsilon _{\theta }=20\varepsilon _{\phi }^{(1)} \)
and \( \varepsilon _{\phi }^{(3)}=0.5\varepsilon _{\phi }^{(1)} \)
are defined in terms of \( \varepsilon _{\phi }^{(1)} \), which can
be varied to test the effect of changing the strength of the local
interactions, while \( \varepsilon _{\rm native} \) can also be varied (see
below) in order to test the effect of changing the strength of the
non-local interactions. The energy of the system is thus controlled
by three parameters \( \varepsilon  \), \( \varepsilon _{\rm native} \)
and \( \varepsilon _{\phi }^{(1)} \). All interaction parameters
are taken to be temperature independent in the present study.

Apart from the variable parameters, this energy function differs from
the similar function used in \cite{Kay03a,Cle00,Kog01}
in two further important ways, 
as follows. (1) For \( r^{ij}/r_{0}^{ij}<\sqrt{5/6} \),
we set \( \varepsilon _{\rm native}=\varepsilon  \), while for \(
r^{ij}/r_{0}^{ij}\geq \sqrt{5/6} \),
we set \( \varepsilon _{\rm native}=\varepsilon _{a} \). Then we can
vary the native interaction parameter \( \varepsilon _{a} \), in
order to test the effect of changing the strength of the non-bonded
attractive interactions between residues, while the short-range repulsive
part of \( V_{\rm native} \) maintains the self-avoidance of the chain.
(2) The value \( q \), which is the smallest number of places along
the chain by which two residues can be separated if they are to interact
by \( V_{\rm nonnative} \), can be set either to \( q=4 \) (in order
to eliminate any double counting of local interactions, in situations
where \( \varepsilon ^{(1)}_{\phi } \) is not being varied) or to
\( q=2 \) (in order to allow \( \varepsilon _{\phi }^{(1)} \) to
decrease without compromising the self-avoidance of the chain).

The equation of motion of each residue is
\begin{equation}
\label{fdg}
m\frac{\partial v^{i}(t)}{\partial t}=F_{\rm conf}^{i}(t)-m\gamma v^{i}(t)+\eta
^{i}(t),
\end{equation}
where \( m \) is the mass of a residue (set to unity), \( \gamma  \)
is the coefficient of friction, \( t \) is time, and \( v^{i}(t) \),
\( F_{\rm conf}^{i}(t) \) and \( \eta ^{i}(t) \) represent each of the
three components of the velocity, conformational force and random
force, respectively \cite{Vei97}. 
The random force is given by\begin{equation}
\label{fdh}
\eta ^{i}(t)=\sqrt{\frac{2m\gamma k_{B}T}{\delta t}}\xi ^{i},
\end{equation}
where \( \delta t \) is the integration time step and \( \xi ^{i} \)
is a random variable taken from a Gaussian distribution with zero
mean and unit variance. The most appropriate time scale can be estimated
\cite{Vei97} by \( \tau =\sqrt{m_{0}a_{0}^{2}/\varepsilon _{0}} \),
where \( m_{0} \), \( a_{0} \) and \( \varepsilon _{0} \) are the
mass, length and energy scales, respectively. We set \( m_{0}=m=1 \),
\( \varepsilon _{0}=\varepsilon =1 \) and \( a_{0}=4 \)~\AA
(the latter is approximately the length of a virtual bond between two
residues and is also the range \( r_{\rm rep} \) of the repulsive interaction),
and so \( \tau =4 \). 
We define the integration time step \( \delta t=0.005\tau  \)
and the coefficient of friction \( \gamma =0.05\tau ^{-1} \) in terms
of this time scale. The velocity-verlet algorithm 
\cite{Kay03a,Vei97,All87} is used to integrate the equations of motion.

\subsection{Lattice models}

The lattice models considered here are 27mers with a maximally
compact native (ground-state) conformation. Details of the models have 
been described elsewhere \cite{Kay03c,Kay03d}. 
We compare three native-centric 
interaction scenarios which have varying degrees, and different mechanisms, of 
thermodynamic cooperativity. As an example of a particular native conformation
to which these three scenarios can be applied, we choose the one in 
Ref.~\cite{Kay03d} with 
relative contact order $0.410$. In scenario (i), which corresponds to the 
common G\=o model, the native contact interactions are pairwise additive. 
In scenario (ii), we add an extra favorable energy $E_{\rm gs}$ for the 
native structure as a whole (as defined by equation~5 in Ref.~\cite{Kay03c}).
Scenario (iii) introduces, in place of the extra favorable energy, a coupling 
between the strength of the contact interaction and the local geometry: two 
residues which are in contact in the native state will interact strongly 
only when the local geometries of the protein chain around the residues are 
the same as those in the native state, as described in 
Ref.~\cite{Kay03d}. We 
characterize this mechanism as local-nonlocal coupling or ``a cooperative 
interplay between favorable nonlocal interactions and local conformational 
preferences'' \cite{Kay03d}. In this scenario, the strength of the native
contact 
interaction is reduced by an 
attenuation factor \( a \) when the local geometry is nonnative. The common
(uncoupled) G\=o model is 
equivalent to \( a=1 \) (no attenuation), while \( a=0 \) implies complete 
coupling; $a=0$ is used here. Under scenarios (i) and (iii), the native 
state has an energy of $-28$ units, while the extra favorable native
energy in scenario (ii) changes the energy of the native state to $-42$ units.
Standard Monte Carlo methods are used for conformational sampling 
\cite{Kay03c,Kay03d}.
The permitted chain moves are end flips, corner flips, crankshafts
and rigid rotations. Each attempted move is counted as one simulation time
step, irrespective of whether the move is accepted by the Metropolis
criterion.
\\

\section{Thermodynamics}

All simulations are performed at constant temperature, with no
explicit consideration of pressure. This is because the focus of
the present study is protein behavior under atmospheric pressure,
and the contribution of a \( PV \) term to protein 
energetics is small under these conditions \cite{Cha00}.  
Therefore, for our present
purposes, we can
consider the Helmholtz and Gibbs free energies to be equivalent. 

\subsection{Calculation of the heat capacity}

The specific heat capacity \( C_{V}(T) \) of the model protein is 
given by the standard relation
\begin{equation}
\label{ffo}
C_{V}(T)=\frac{1}{k_{B}T^{2}}\left[ \left\langle E^{2}(T)
\right\rangle -\left\langle E(T) \right\rangle^{2}\right] ,
\end{equation}
where \( k_{B} T \) is the Boltzmann constant multiplied by the absolute
temperature, 
\( \left\langle X(T) \right\rangle \) denotes the Boltzmann average 
of quantity \( X \) at temperature \( T \), and the total energy
\( E \) is the sum \( V+E_{K} \) of potential and kinetic energies.
We compute the averages by standard histogram sampling techniques 
\cite{Kay03a,Cle00}.

In lattice studies, the kinetic energy \( E_{K} \) is not treated. 
Therefore, \( E \) in Eq.~(\ref{ffo}) has traditionally been taken, in protein
modeling, to be the
potential energy term \( V \). This procedure 
has often been extended to continuum model studies, in Ref.~\cite{Kay03a} 
for example, 
although \( E_{K} \) is accessible and well-defined in off-lattice models. 
However, $\langle E^2\rangle$ $-$ $\langle E\rangle^2$ $\ne$
$\langle V^2\rangle$ $-$ $\langle V\rangle^2$ in general. The two quantities
would be equal if $E_K$ were a constant, but that would be unphysical.
In this study, we have calculated \( C_{V}(T) \) using Eq.~(\ref{ffo}) both 
with $E=V+E_{K}$ and with the substitution $E\rightarrow V$. 
The results are not identical: an example is given in Fig.~1.

 Fig.~1 shows that the difference between heat capacity values obtained 
using the two methods is small around the transition midpoint \( T_{m} \).
This is because any energy added to the system during the unfolding
transition contributes mostly to the potential rather than to the kinetic energy. 
The difference is less negligible for the ``shoulders'' on either side 
of the heat capacity peak. At very low temperatures, including the kinetic 
energy contribution can lead to a smaller heat capacity, because the 
molecule at this temperature is in a relatively fixed state: nearly all of 
the kinetic energy is accounted for by the oscillation of pairs of residues 
about the minima of their mutual (bonded or non-bonded) interaction 
energies. The potential energy and the kinetic energy associated with 
these oscillations both fluctuate, but their sum fluctuates much less, and 
so the fluctuations in total energy are smaller than the fluctuations in 
potential energy, with the result that the calculated heat capacity is 
smaller when \( E_{K} \) is taken into account. Overall, Fig.~1 indicates
that while the difference between the heat capacities calculated using the two 
different methods is not negligible, it is not drastic. Probably this
is because \( E_{K} \), while not invariant, fluctuates much less than 
\( V \). For this reason, we do not expect conclusions drawn from previous 
calculations of heat capacities \cite{Kay03a}, which used \( V \), to be 
changed 
greatly by calculations using \( E \). Nonetheless, we do expect a proper 
account of the kinetic contributions to protein heat capacities
to be important in addressing the contribution of bond vector motions
to the heat capacity \cite{Yan97}. 

Since the \( PV \) term is neglected in the present formulation, 
\( C_{V}(T) \) is effectively equal to \( C_{P}(T) \), which is generally 
measured by calorimetry (and which can be expressed in a form similar to 
Eq.~(\ref{ffo}), but with the enthalpy \( H \) taking the place of the 
energy \( E \) \cite{Cha00,Kay00a}). 
Therefore, we may refer to the quantity computed 
using Eq.~(\ref{ffo}) simply as heat capacity. All subsequent heat capacity 
curves shown in this article for the continuum models are obtained using 
the total energy \( E=V+E_K \).

\subsection{The free energy}

The Helmholtz free energy of the model system is \( F(T) = -k_{B}T\ln Z(T) \), 
where \( Z(T) \) is the partition function at temperature \( T\). It 
follows that, in the vicinity of the simulation temperature \( T_{\rm sim} \),
the Helmholtz free energy of the model system at temperature \( T \), relative to
its value at the simulation temperature, may be approximated using the formula:
\begin{equation}
\label{ffna}
\frac{\Delta F(T)}{k_{B}T}=
\frac{F(T)}{k_{B}T}-\frac{F(T_{\rm sim})}{k_{B}T_{\rm sim}}
=-\ln \left\{ \sum_{i}p(E_{i};T_{\rm sim})\exp 
\left( E_{i}\left[ \frac{1}{k_{B}T_{\rm sim}}-\frac{1}{k_{B}T}\right] 
\right) \right\}, 
\end{equation}
where the sum is performed over sets of microstates in different energy
ranges $E_i$. \( p(E_{i};T_{\rm sim}) \) is the probability density at 
the simulation temperature, and is estimated directly from the Langevin 
dynamics simulations.

The inset of Fig.~1 provides an example of \( \Delta F(T) \), showing that 
the gradient of the free energy with 
respect to $T$ changes rather abruptly around the transition 
temperature \( T_{m} \) (vertical dotted lines). The transition
temperature \( T_{m} \) corresponds to the temperature at the peak of 
the heat capacity curve, which was denoted by \( T_{\rm max} \) in 
Ref.~\cite{Kay00a}. 
Apparently, below \( T_{m} \), the system spends most time in states 
in the vicinity of the bottom of the native basin, and so the 
changes in \( F \) with respect to temperature are dominated by the 
behavior of these states. However, as the temperature 
increases past \( T_{m} \), the system, and therefore the rate of 
change of \( F(T) \), starts to be dominated by states near
the bottom of the denatured basin. As a result, the
gradient of \( F(T) \) changes rather suddenly at \( T_{m} \). 
While the \( F(T) \) gradient could never be discontinuous because the
model system is finite, the kink at \( T_{m} \) does indicate
that the transition is two-state-like, and therefore that it is 
similar to a first order phase transition.

\subsection{Thermodynamic cooperativity}

The presence of a peak in the heat capacity at a transition temperature
\( T_{m} \), as in Fig.~1, indicates that the folding/unfolding transition
possesses a degree of thermodynamic cooperativity. As our group has
argued, the degree of thermodynamic cooperativity in protein models 
can be quantified by the ratio 
\( \kappa_{2}=\Delta H_{\rm vH}/\Delta H_{\rm cal} \) of the van't Hoff 
enthalpy \( \Delta H_{\rm vH} \) to the calorimetric enthalpy 
\( \Delta H_{\rm cal} \) of the transition. This ratio is closely
related to that determined experimentally by differential scanning
calorimetry \cite{Pri74}. In model studies, the calorimetric enthalpy 
$\Delta H_{\rm cal}$ may be determined from an integral of the heat 
capacity across the transition region,
\begin{equation}
\label{ffua}
\Delta H_{\rm cal}=\int dT C_{P}(T),
\end{equation}
while the van't Hoff enthalpy is equal to twice the maximum standard
deviation of the enthalpy distribution at the transition midpoint,
\begin{equation}
\label{ffub}
\Delta H_{\rm vH}=2\sqrt{k_{B}T_{m}^{2}C_{P}^{\rm max}},
\end{equation}
where \( C_{P}^{\rm max} \) is the peak value of the heat capacity. We
have followed standard usage in this section by expressing \( \kappa _{2} \)
in terms of \( H \) and \( C_{P} \). However, as mentioned in the
previous section, simulations produce values for \( E \) and \( C_{V} \),
which for the present application are essentially equivalent 
to \( H \) and \( C_{P} \). 

As has been pointed out \cite{Kay00a}, comparison of simulation heat capacity 
scans to experiment is often complicated by the fact that the heat capacity
tails which we observe in simulations would, if they occurred in a real 
system, be swamped by the solvent contribution and ignored by the
common procedure of using empirical baseline subtraction to calculate
$\Delta H_{\rm vH}/\Delta H_{\rm cal}$. In other words, tail contributions 
that arise from conformational transitions may be masked by solvent 
contributions in real data analysis \cite{Kay00a,Kay02}. Therefore, for
completeness,
we also perform empirical baseline subtractions on our simulated heat 
capacity scans, producing a revised ratio \( \kappa^{({\rm s})}_{2} \) 
(defined in Ref.~\cite{Kay00a}) to facilitate comparison with experiment.

\subsection{Radius of gyration}

The radius of gyration \( R_g \) of a particular conformation of the
protein is an indicator of its compactness. It is defined by
\begin{equation}
\label{ffv}
R_g^{2}=\frac{1}{N}\sum _{i=1}^{N}\left| r_{i}-\left\langle r\right\rangle \right| ^{2},
\end{equation}
where \textbf{\( N \)} is the number of residues, \( r_{i} \)
is the position of the \( i \)th residue, and $\langle r\rangle$ is the
average position (centroid) of the given conformation. The Boltzmann
average \( \left\langle R_g\right\rangle =\left\langle \sqrt{R_g^{2}}
\right\rangle \) over a 
given conformational ensemble is obtained by standard histogram techniques.
Two-state-like behavior requires a steplike sigmoidal change in 
\( \left\langle R_g\right\rangle  \) upon denaturation at \( T_{m} \), with
little postdenaturational expansion of the chain \cite{Kay00a}.

\section{Results and discussion}

To study the effect of local vs. nonlocal interactions, we first
vary the strength $\varepsilon_\phi^{(1)}$ of the local interactions while 
keeping the strength of the nonlocal interactions fixed in the continuum CI2 
construct (Figs.~2, 3). The heat capacity scans, for four scenarios (four
models) with different values for $\varepsilon_\phi^{(1)}$, are shown in Fig.~2.
They 
all exhibit a fairly sharp peak except for the
model with $\varepsilon_\phi^{(1)}=0.25$. The
heat capacity peak signifies substantial heat absorption within 
a narrow temperature range at the folding/unfolding transition. The absorbed
energy propels the chain from its low-energy folded conformations (native 
ensemble) to its high-energy unfolded conformations (denatured ensemble). 
However, the mere existence of a relatively sharp peak in the heat capacity 
function does not necessarily mean that the transition is as cooperative as 
those observed in small single-domain proteins. Coil-globule transitions in
homopolymers 
are not two-state-like, but their calorimetric heat capacity scans can have 
very sharp peaks \cite{Tik94}. A more quantitative measure of 
thermodynamic cooperativity is the traditional calorimetric two-state 
criterion (see above), which has emerged recently as a powerful modeling 
tool \cite{Cha00,Kay00a,Kay00b,Kay02,Kay03a,Cri02,Jan02,Pok03,Cle03}. 
Ratios of van't Hoff to calorimetric enthalpy were 
calculated for the four models as described above; the ranges of the temperature 
integrations used in the determination of $\Delta H_{\rm cal}$ [Eq.~(\ref{ffua})]
were taken to be equal to the ranges shown in Fig.~2.

The inset of Fig.~2 shows that the $\Delta H_{\rm vH}/\Delta H_{\rm cal}$ 
ratio (diamonds) of these models is only weakly dependent on
$\varepsilon_{\phi}^{(1)}$ over an extended range of 
$\varepsilon_\phi^{(1)}$ values, but that the ratio is significantly smaller
when the local interactions are 
substantially weaker, at $\varepsilon_\phi^{(1)}=0.25$, than the nonlocal 
interactions. As discussed above, quantitative comparisons between 
simulated and experimental $\Delta H_{\rm vH}/\Delta H_{\rm cal}$ values 
require the introduction of model calorimetric baselines \cite{Kay00a} 
similar to 
those employed in the interpretation of experimental data. Traditionally, 
experimental baselines are designed to remove solvation contributions 
(temperature-dependent effective interactions), in order to extract the heat
capacity effects 
associated with the folding/unfolding transition itself \cite{Pri74}. 
The present 
models do not contain temperature-dependent interactions. Therefore, the 
heat capacity contributions eliminated by the model calorimetric 
baselines in Fig.~2 can only originate from vibrational motions and 
conformational transitions. Increasingly, it is being recognized 
\cite{Kay00a,Yan97,Dra02} 
that similar heat capacity contributions from bond vector motions and 
more collective conformational transitions might also be ``hidden'' below 
traditional baselines constructed for analyzing experimental calorimetric data,
although the magnitude of such contributions needs to be elucidated.
 The three models in Fig.~2 that are relatively more cooperative
(with higher $\kappa_2$ values) all have modified 
$\Delta H_{\rm vH}/\Delta H_{\rm cal}$ values ($\kappa_2^{({\rm s})}$,
circles in the inset), after empirical baseline subtractions, that are very 
close to unity. (We note that the recent determination of 
$\Delta H_{\rm vH}/\Delta H_{\rm cal}$ values in an all-atom G\=o model
\cite{Cle03}
involved baseline subtractions as well: c.f. Fig.~8 of Ref.~\cite{Cle03}.) 

However, a protein chain model's ability to attain a near-unity 
$\Delta H_{\rm vH}/\Delta H_{\rm cal}$ ratio after baseline subtractions 
does not by itself imply that its thermodynamic behavior is similar to that of 
real, small single-domain proteins \cite{Kay00a,Kay00b}. This is because the 
heat capacity contributions discarded by certain baselines can actually be
symptoms of significant deviations from two-state-like behavior. It has been
recognized \cite{Kay00a} that, to clarify this situation, we can use the
behavior of the 
average radius of gyration $\langle R_g \rangle$ of a protein chain model as
an
additional evaluation criterion for the model's thermodynamic cooperativity.
Small angle X-ray scattering (SAXS) experiments have demonstrated that the
average 
radius of gyration $\langle R_g \rangle$ of several small single-domain proteins
behaves 
in an apparently two-state manner \cite{Sos92,Hag98,Mil02}, 
showing very little 
postdenaturational ($T>T_{m}$) expansion of the chain outside the 
transition regime that corresponds to the region of the heat capacity
peak. We require chain models 
of small single-domain proteins to exhibit similar behavior \cite{Kay00a}. 
Now, to further assess the four models in Fig.~2 with different 
local interaction strengths, we calculate their average radii of
gyration as a function of temperature (Fig.~3). To 
ensure adequate sampling, $\langle R_g\rangle$ for each model is obtained 
from three different simulation temperatures; the results are thus 
displayed as three discontinuous curves. Despite some minor discrepancies
(owing to sampling uncertainties) between parts of the $\langle R_g\rangle$ 
function deduced from different simulation temperatures for the 
$\varepsilon_\phi^{(1)}=0.25$ case, the general trend in Fig.~3 is very
clear. Models with weaker local interactions are less cooperative 
in that their $\langle R_g\rangle$ curves show more postdenaturational increase
than do those of models having stronger local interactions. For instance, 
the $\langle R_{g} \rangle$ of the $\varepsilon_{\phi}^{(1)}=0.25$ model
increases by $\approx 3.0$~\AA~ between $T \approx 0.82$ (the end of the
transition region) and $T \approx 1.11$. In contrast, a similar
temperature increase for the 
$\varepsilon_\phi^{(1)}=1.00$ model from $T\approx 1.11$ (the end of the
transition region) to $T\approx 1.40$ leads to
an increase of only $\approx 1.6~$\AA~ in $\langle R_g\rangle$.
These observations indicate that two-state-like thermodynamic cooperativity
cannot be achieved if the local conformational propensities of a protein 
are much weaker than the favorable nonlocal interactions. This confirms a 
similar conclusion which was derived recently from a more limited study of a 
``contact dominant model'' \cite{Kay03a}.

We next extend our analysis by applying the same computational
procedure to varying the strength $\varepsilon_a$ of the favorable 
nonlocal interactions while keeping the strength of the local 
interactions fixed. Consistent with the seminal study of G\=o and Taketomi 
\cite{Go_78}, Figs.~4 and 5 show that variations in nonlocal $\varepsilon_a$
have a more prominent effect on thermodynamic cooperativity than 
variations in local $\varepsilon_\phi^{(1)}$. 
While the peak heat capacity values for the three models in Fig.~2 with 
$\varepsilon_\phi^{(1)}\ge 0.5$ are similar, the peak heat capacity
values for the three models in Fig.~4 with $\varepsilon_a\ge 0.5$
show a significant monotonic increase with $\varepsilon_a$. In addition,
for the $\varepsilon_a=0.5$ model in Fig.~4, the difference
between unity and the $\Delta H_{\rm vH}/\Delta H_{\rm cal}$ ratio 
after baseline subtraction is not negligible ($\kappa_2^{({\rm s})}=0.91$).
Despite these differences, the trends in Figs.~4,~5 are in large measure 
similar to those in Figs.~2,~3. In particular, Fig.~4 shows that the model with
$\varepsilon_a=0.25$, like the $\varepsilon_\phi^{(1)}=0.25$ case in Fig.~2,
has a significantly
lower $\Delta H_{\rm vH}/\Delta H_{\rm cal}$ ratio than the other
three models considered in the same figure. The 
$\langle R_g\rangle$ data in Fig.~5 shows that thermodynamic cooperativity 
increases with $\varepsilon_a$, as manifested in a smaller
amount of postdenaturational
conformational expansion with increasing $\varepsilon_a$; this is comparable
to the effect of increasing $\varepsilon_{\phi}^{(1)}$ in Fig.~3.
Taken together, 
the results in Figs.~2--5 suggest that a high degree of thermodynamic 
cooperativity, similar to that in real, small single-domain proteins,
requires both strong local and strong nonlocal interactions. Apparently,
a high degree of thermodynamic cooperativity is {\it incompatible} with either 
a much weakened local conformational preference relative to the favorable 
nonlocal interactions ($\varepsilon_\phi^{(1)}\ll \varepsilon_a$) or much 
weakened favorable nonlocal interactions relative to the local conformational 
preference ($\varepsilon_a\ll \varepsilon_\phi^{(1)}$).

Although three-dimensional G\=o-like models with strong local 
and nonlocal interactions appear to satisfy the thermodynamic criterion 
of calorimetric two-state cooperativity, it has recently been noted that 
they are unable to produce simple two-state folding/unfolding kinetics 
\cite{Kay02,Kay03a}.
This is because the thermodynamic cooperativity of these models is not
sufficiently high. As a result, and in spite of the native-centric nature
of the common pairwise additive G\=o-like interactions, kinetic trapping 
becomes significant under strongly native conditions, leading to folding 
rate slow-downs and chevron rollovers \cite{Kay03b}. More recent lattice 
model 
investigations indicate that simple two-state folding/unfolding kinetics 
require a high degree of thermodynamic cooperativity that may be
characterized as ``near-Levinthal'' \cite{Kay03c}, necessitating many-body 
interactions beyond those postulated by the common G\=o model 
\cite{Kay03c,Kay03d}.

In view of this recent development, and to facilitate the construction and
investigation of continuum models that incorporate these new ideas,
it is instructive to compare in more detail the thermodynamic behavior of 
the common lattice G\=o construct (with only pairwise additive contact
energies) with the behavior of models which have many-body interactions and
enhanced cooperativity. We also wish to investigate whether results obtained from
lattice models supply additional support for the conclusions which we have derived
from our continuum model results.
To this end, Figs.~6--8 compare three 27mer lattice models. 

Because of their intrinsic restrictions on conformational possibilities,
it is more straightforward to construct cooperative lattice models 
than to construct off-lattice continuum models that are similarly cooperative.
Recently, using evidence from kinetic simulations of chevron plots, our group has
proposed 
that a $\Delta H_{\rm vH}/\Delta H_{\rm cal}$ ratio of $\kappa_2>0.9$ before 
baseline subtractions [as for models (ii) and (iii) in Fig.~6a] is likely to be 
required in order for a lattice protein chain model to produce chevron 
plots with linear regimes similar in extent to those observed for real, small
single-domain 
proteins \cite{Kay03c}. 
However, this numerical criterion is not readily generalizable
to off-lattice continuum models. This is because the heat capacity effects
of bond vibrations and kinetic energy have to be taken into account in
continuum models, whereas these effects are absent in lattice models. Thus, in
the characterization of a model's thermodynamic cooperativity, more detailed
information concerning, for example, the behavior of the average radius of
gyration, has to be relied upon more heavily for continuous models (see above)
than for lattice models.

 For the three lattice models studied here, the $\langle R_g\rangle$ plots 
in Fig.~6b show that, while the postdenaturational conformational expansion 
of the common lattice G\=o construct (solid curve in Fig.~6b) is considerably 
milder than that of its continuum counterpart (solid curves in Figs.~3, 5),
the more cooperative lattice models with many-body interactions 
exhibit much less (dotted curve in Fig.~6b) or nearly non-existent
(dashed curve in Fig.~6b) postdenaturational conformational expansion.
The fluctuations in $E$ and $R_g$ near the transition midpoint, shown in 
 Figs.~7 and 8, indicate further that the transitions between the native 
and denatured ensembles are sharper and more two-state-like for the more
cooperative models (ii) and (iii) with many-body interactions [parts (b) and (c) 
of Figs.~7, 8] than for the common G\=o model [parts (a) of Figs.~7, 
8]. The corresponding fluctuations in the fractional number of native 
contacts $Q$ \cite{Kay03c,Kay03d} (data not shown) were also found 
to exhibit a trend 
very similar to that of the energy fluctuations in Fig.~7.

The lattice model results, shown in Figs.~6--8, are compatible with the
conclusion, reached above on the basis of continuum model results, that both
local and nonlocal interactions are important for thermodynamic cooperativity.
The common lattice G\=o model of scenario (i) includes only nonlocal interactions,
analogous to the interactions encoded by $V_{\rm native}$ in the off-lattice
model. Scenario (iii), which takes account also of the local geometry of the
chain, displays greater cooperativity than scenario (i).

Higher resolution data such as that in Figs.~7 and 8 opens up future 
avenues for the assessment of different mechanisms of cooperativity using
comparisons between model
predictions and experimental measurements of, for example, conformational
sizes and fluctuations \cite{Cho02,Shi02,Gol03}. 
It is noteworthy that in the model (ii) 
scenario, with an extra favorable energy for the native structure as a whole,
the native ensemble does not exhibit much energetic or conformational
variation (horizontal line segments at low $E$ and low $R_g$ values in Figs.~7b 
and 8b). On the other hand, in the model (iii) scenario with local-nonlocal
coupling, there is considerable variation in the native ensemble
(c.f. low $E$
and low $R_g$ fluctuations in Figs.~7c and 8c). Yet the variation
in the denatured ensemble is smaller in model (iii) than in model (ii)
(c.f. high $E$ and high $R_g$ fluctuations in parts (b) and (c) of 
 Figs.~7, 8), resulting in a more two-state-like average $\langle R_g\rangle$
transition for model (iii) than for model (ii), manifested in a 
near-immediate postdenaturational ($T>T_{m}$) saturation of the dashed 
curve for model (iii) in Fig.~6b compared to a more gradual 
postdenaturational saturation of the dotted curve for model (ii) in the 
same figure. All these differences in conformational properties are in 
principle detectable through experiments on real proteins. Hence, future 
experimental efforts along the lines suggested here would help to verify 
or falsify different proposed scenarios and interaction mechanisms 
\cite{Kay03c,Kay03d,Jew03,Cha03} 
in the endeavor to decipher the physical origins of 
cooperativity in real proteins. 

\section{Conclusions}

The present study suggests strongly that both local conformational
preferences and favorable nonlocal interactions are crucial to
protein thermodynamic cooperativity. 
This result points to a useful constraint on simplified models of protein 
molecules: they should take account both of local and of nonlocal
interactions. 

As emphasized above, the scope 
of the present study is limited. Only native-centric interaction 
schemes are considered; and here we have elected only one particular 
physically plausible way to classify energy contributions into ``local'' 
and ``nonlocal'' terms in the continuum models. 
In addition, by using a native-centric interaction scheme both for local and
for nonlocal interactions, we have avoided the possibility of energetic 
frustration, which might be significant if a more realistic interaction 
scheme were used \cite{Pac03,HG03}. To further elucidate the 
answers to the questions we have posed, much remains to be investigated.

Nonetheless,
our results show clearly that a high degree of thermodynamic cooperativity is
compatible neither with a much weakened local interaction
nor with a much weakened nonlocal interaction, indicating that both
local and nonlocal interactions are important components in protein
energetics \cite{Uve02}. This finding is consistent with the notion that
a cooperative interplay between local and nonlocal interactions 
\cite{Cha00,Kay00b,Kay02,Kay03c,Kay03d} 
is a critical ingredient underlying the apparent simple two-state
cooperativity of real, small single-domain proteins. 

With regard to G\=o-like
native-centric modeling (see, e.g., discussion in 
Refs.~\cite{Kay03a,Cle00,Mic99,Plo01,Cie02,Cie03}), 
we observe that significant differences in model 
predictions can result from different G\=o-like interaction schemes,
even though all of the schemes are designed to bias the chain towards the 
same native structure. This underscores our point that requiring a
consistent account of cooperativity can be a more productive approach
to protein modeling than simply designing a model heteropolymer to 
fold to a target structure \cite{Kay03a}. 
In this context, the present coarse-grained 
representations constitute only a first step in the understanding of protein 
cooperativity. Ultimately, atomistic origins of local and nonlocal
interactions such as sidechain packing \cite{Kli98,Li_01,Fav02,Zho02} 
must be taken into
account in an effort to provide the necessary physical underpinning 
for the cooperative mechanisms proposed here.

\section*{Acknowledgments}
The research reported here was partially supported by the Canadian 
Institutes of Health Research (CIHR grant no. MOP-15323), PENCE, 
a Premier's Research Excellence Award from the Province of Ontario, and 
the Ontario Centre for Genomic Computing at the Hospital for Sick 
Children in Toronto. H. S. C. holds a Canada Research Chair in Biochemistry.

\vfill\eject

\vfill\eject

\noindent
{\large\bf Figure Captions}

\noindent
FIGURE 1

Heat capacity as a function of temperature, for one particular set
of interaction parameters. Solid curve: heat capacity calculated using
the total energy; dashed curve: heat capacity calculated 
using only the potential energy. 
Parameters are \( \varepsilon _{\phi }^{(1)}=1.00 \),
\( \varepsilon _{a}=1.00 \), \( q=4 \); 
simulation temperature \( T_{\rm sim}=1.02 \).
Inset: free energy as a function of temperature, for the same model,
showing a sharp change in gradient around \( T_{m} \).
The vertical dotted lines in the figure and the inset mark the transition
midpoint temperature \( T_{m} \).
\\

\noindent
FIGURE 2

Heat capacity as a function of temperature, for varying local interaction
energy \( \varepsilon _{\phi }^{(1)} \). Other parameters 
\( \varepsilon_{a}=1.00 \) and \( q=2 \) are fixed. ({\it From left to right})
dotted curve: \( \varepsilon _{\phi }^{(1)}=0.25 \), \( T_{m}=0.74 \); 
short dashed curve: \( \varepsilon _{\phi }^{(1)}=0.50 \), \( T_{m}=0.84 \);
long dashed curve: \( \varepsilon _{\phi }^{(1)}=0.75 \), \( T_{m}=0.94 \);
solid curve: \( \varepsilon _{\phi }^{(1)}=1.00 \), \( T_{m}=1.03 \).
These scans are obtained by histogram techniques from simulations performed
at $T_{\rm sim}=$ $0.73$, $0.84$, $0.94$, and $1.03$ respectively.
The $\Delta H_{\rm vH}/\Delta H_{\rm cal}$ cooperativity coefficients 
\( \kappa_2  \) without baseline subtractions are $0.33$, $0.43$, $0.44$, 
and $0.44$ respectively. Modified cooperativity coefficients 
\( \kappa_2^{({\rm s})} \) after subtraction of the baselines
(indicated by thin lines in the figure) for $\varepsilon_{\phi }^{(1)}=$
$0.50$, $0.75$, and $1.00$ are $0.97$, $0.98$, and $0.99$ respectively.
No value for \( \kappa_2^{({\rm s})} \) was calculated for 
\( \varepsilon_{\phi }^{(1)}=0.25 \) because the shape of its heat capacity 
curve does not suggest any clear choice of baselines that are intuitively
more reasonable than others. The inset shows \( \kappa_2  \) (diamonds) and 
\( \kappa_2^{({\rm s})} \) (circles) as functions of 
\( \varepsilon_{\phi }^{(1)} \).
\\

\noindent
FIGURE 3

Average radius of gyration as a function of temperature, for varying
local interaction energy \( \varepsilon _{\phi }^{(1)} \); other
parameters \( \varepsilon _{a}=1.00 \) and \( q=2 \) are fixed, as in Fig.~2.
The correspondence between line styles and \( \varepsilon _{\phi }^{(1)} \)
values is identical to that in Fig.~2.
For each value of \( \varepsilon _{\phi }^{(1)} \), simulations were
performed at three different values of $T_{\rm sim}$ to ensure adequate
sampling
across the entire temperature range shown. ({\it From left to right})
for \( \varepsilon _{\phi }^{(1)}=0.25 \) (dotted curves),
\( T_{\rm sim}=0.73 \), $0.93$, $1.13$;
for \( \varepsilon _{\phi }^{(1)}=0.50 \) (short dashed curves),
\( T_{\rm sim}=0.84, \) $1.04$, $1.24$; 
for \( \varepsilon _{\phi }^{(1)}=0.75 \) (long dashed curves),
\( T_{\rm sim}=0.94, \) $1.14$, $1.34$; and
for \( \varepsilon _{\phi }^{(1)}=1.00 \) (solid curves),
\( T_{\rm sim}=1.03 \), $1.23$, and $1.43$.
\\

\noindent
FIGURE 4

Heat capacity as a function of temperature, for varying nonlocal interaction
energy \( \varepsilon _{a} \). Other parameters 
\( \varepsilon _{\phi }^{(1)}=1.00 \) and \( q=4 \) are fixed. 
({\it From left to right})
dotted curve: \( \varepsilon _{a}=0.25 \), \( T_{m}=0.43 \);
short dashed curve: \( \varepsilon _{a}=0.50 \), \( T_{m}=0.65 \); 
long dashed curve: \( \varepsilon _{a}=0.75 \), \( T_{m}=0.84 \); 
solid curve: \( \varepsilon _{a}=1.00 \), \( T_{m}=1.02 \).
These scans are obtained by histogram techniques from simulations performed
at $T_{\rm sim}=$ $0.42$, $0.64$, $0.84$, and $1.02$ respectively.
The $\Delta H_{\rm vH}/\Delta H_{\rm cal}$ cooperativity coefficients 
\( \kappa_2  \) without baseline subtractions are $0.28$, $0.40$, $0.44$, 
and $0.46$ respectively. Modified cooperativity coefficients 
\( \kappa_2^{({\rm s})} \) after subtraction of the baselines
(indicated by thin lines in the figure) for $\varepsilon _{a}=$
$0.50$, $0.75$, and $1.00$ are $0.91$, $1.00$, and $0.99$ respectively.
No value for \( \kappa_2^{({\rm s})} \) was calculated for 
\( \varepsilon _{a}=0.25 \) for the same reason that no 
\( \kappa_2^{({\rm s})} \) was provided for 
\( \varepsilon_{\phi }^{(1)}=0.25 \) in Fig.~2.
The inset shows \( \kappa_2  \) (diamonds) and \( \kappa_2^{({\rm s})} \) 
(circles) as functions of \( \varepsilon _{a} \).
\\

\noindent
FIGURE 5

Average radius of gyration as a function of temperature, for varying
nonlocal interaction energy \( \varepsilon _{a} \). Other parameters
\( \varepsilon _{\phi }^{(1)}=1.00 \) and \( q=4 \) are fixed, as in Fig.~4.
The correspondence between line styles and \( \varepsilon _{a} \)
values is identical to that in Fig.~4.
For each value of \( \varepsilon _{a} \), simulations were performed
at three different values of $T_{\rm sim}$ to ensure adequate sampling across
the whole range shown.  ({\it From left to right}) 
for \( \varepsilon _{a}=0.25 \) (dotted curves), \( T_{\rm sim}=0.42 \),
$0.62$, $0.82$;
for \( \varepsilon _{a}=0.50 \) (short dashed curves), \( T_{\rm sim}=0.64 \),
$0.84$, $1.04$; 
for \( \varepsilon _{a}=0.75 \) (long dashed curves), \( T_{\rm sim}=0.84, \)
$1.04$, $1.24$; and 
for \( \varepsilon _{a}=1.00 \) (solid curves), \( T_{\rm sim}=1.02 \),
$1.22$, and $1.42$.
\\

\noindent
FIGURE 6

Thermodynamic cooperativities of the three representative 27mer
lattice models described in the text. Model definitions and simulation 
details are given in Refs.~[29,30]. Heat capacity (a) and average 
radius of gyration (b) are determined by histogram techniques based 
upon Monte Carlo sampling performed at $T_{\rm sim}=T_{m}$.
(a) Heat capacity of (i) the common G\=o model with pairwise additive 
contact energy (solid curve, {\it left}); (ii) the model that assigns an 
extra favorable energy to the native structure as a whole (dotted curve, 
{\it right}); and (iii) the model with local-nonlocal coupling (dashed curve, 
{\it middle}). The transition temperatures
for the models are $T_{m}=$ $0.701$ (i), $1.13$ (ii), and $0.755$ (iii).
Their $\Delta H_{\rm vH}/\Delta H_{\rm cal}$ without baseline subtractions
are $\kappa_2=$ $0.86$, $0.98$, and $0.99$, and the corresponding ratios
after subtracting the baselines shown are $\kappa_2^{({\rm s})}=$ 
$1.00$, $1.00$, and $1.00$ respectively. (b) Average radius of gyration 
as a function of model temperature for the three models [represented by 
the same line styles as in (a)].
\\

\noindent
FIGURE 7

Representative trajectories of the three models in Fig.~6 at their
respective transition temperatures.  Variations in the potential energy of models 
(i)--(iii) are shown in (a)--(c) respectively.
\\

\noindent
FIGURE 8

Same as Fig.~7, except that variations in the radius of gyration are shown here.


\begin{thebibliography}{99}

\bibitem{Cha02} Chan HS, Kaya H, Shimizu S. 
In: Jiang T, Xu Y, Zhang MQ, editors. Current Topics in Computational
Molecular Biology. Cambridge, MA: The MIT Press; 2002. p 403--447.
\bibitem{Jac91}Jackson SE, Fersht AR. Biochemistry 1991;30:10428--10435.
\bibitem{Bak00}Baker D. Nature 2000;405:39--42.
\bibitem{Cha00} Chan HS. Proteins 2000;40:543--571.
\bibitem{Kay00a} Kaya H, Chan HS. Proteins 2000;40:637--661 
                      [Erratum: Proteins 2001;43:523].
\bibitem{Kay00b} Kaya H, Chan HS. Phys Rev Lett 2000;85:4823--4826.
\bibitem{Kay02} Kaya H, Chan HS. J Mol Biol 2002;315:899--909.
\bibitem{Kay03a} Kaya H, Chan HS. J Mol Biol 2003;326:911--931.
\bibitem{Tak75} Taketomi H, Ueda Y, G\=o N. 
               Int J Pept Protein Res 1975;7:445--459.
\bibitem{Cle00}
Clementi C, Nymeyer H, Onuchic JN. J Mol Biol 2000;298:937--953.
\bibitem{Mic99} Micheletti C, Banavar JR, Maritan A, Seno F. 
                  Phys Rev Lett 1999;82:3372--3375.
\bibitem{Zho99} Zhou Y, Karplus M. Nature 1999;401:400--403.
\bibitem{Por01} Portman JJ, Takada S, Wolynes PG. 
                 J Chem Phys 2001;114:5082--5096.
\bibitem{Hal97}Haliloglu T, Bahar I, Erman B. 
                  Phys Rev Lett 1997;79:3090--3093.
\bibitem{Kes02} Keskin O, Bahar I, Flatow D, Covell DG, Jernigan RL. 
                Biochemistry 2002;41:491--501.
\bibitem{Isi02} Isin B, Doruker P, Bahar I. Biophys J 2002;82:569--581.
\bibitem{Mic02} Micheletti C, Lattanzi G, Maritan A. 
                 J Mol Biol 2002;321:909--921.
\bibitem{Jac01}Jacobs DJ, Radar AJ, Kuhn LA, Thorpe MF. 
                Proteins 2001;44:150--165.
\bibitem{Kay03b} Kaya H, Chan HS. Phys Rev Lett 2003;90:258104.
\bibitem{Go_78} G\=o N, Taketomi H. Proc Natl Acad Sci USA 1978;75:559--563.
\bibitem{Dil90} Dill KA. Biochemistry 1990;29:7133--7155.
\bibitem{Abk95} Abkevich VI, Gutin AM, Shakhnovich EI. 
J Mol Biol 1995:252:460--471.
\bibitem{Cha98} Chan HS. Nature 1998;392:761--763.
\bibitem{Bal99}Baldwin RL, Rose GD. Trends Biochem Sci 1999;24:26--33.
\bibitem{Por01-2}Portman JJ, Takada S, Wolynes PG. 
                 J Chem Phys 2001;114:5069--5081.
\bibitem{Deg75}de Gennes PG. J Phys Lett (Paris) 1975;36:L55--L57.
\bibitem{Don96}Doniach S, Garel T, Orland H. J Chem Phys 1996;105:1601--1608.
\bibitem{Bas97}Bastolla U, Grassberger P. J Stat Phys 1997;89:1061--1078.
\bibitem{Doy98}Doye JPK, Sear RP, Frenkel D. J Chem Phys 1998;108:2134--2142.
\bibitem{Kog01}Koga N, Takada S. J Mol Biol 2001;313:171--180.
\bibitem{Bra67}Brant DA, Miller WG, Flory PJ. J Mol Biol 1967;23:47--65.
\bibitem{Vei97}Veitshans T, Klimov D, Thirumalai D. Fold Des 1997:2:1--22.
\bibitem{All87}Allen MP, Tildesley DJ. Computer Simulation of Liquids.
Oxford: The Oxford University Press; 1987.
\bibitem{Kay03c}Kaya H, Chan HS, Proteins 2003; in press.
\bibitem{Kay03d}Kaya H, Chan HS, Proteins 2003; in press.
\bibitem{Yan97}Yang D, Mok YK, Forman-Kay JD, Farrow NA, Kay LE.
                 J Mol Biol 1997;272:790--804.
\bibitem{Pri74}Privalov PL, Khechinashvili NN. 
                      J Mol Biol 1974;86:665--684.
\bibitem{Tik94}Tiktopulo EI, Bychkova VE, Ri{\v c}ka J, Ptitsyn OB.
                 Macromolecules 1994;27:2879--2882.
\bibitem{Cri02}Crippen GM, Chhadjer M. J Chem Phys 2002;116:2261--2268.
\bibitem{Jan02}Jang H, Hall, CK, Zhou Y. Biophys J 2002;82:646--659.
\bibitem{Pok03}Pokarowski P, Kolinski A, Skolnick J.
                    Biophys J 2003;84:1518--1526.
\bibitem{Cle03}Clementi C, Garcia AE, Onuchic JN. 
                    J Mol Biol 2003;326:933--954.
\bibitem{Dra02}Dragan AI, Privalov PL. J Mol Biol 2002;321:891--908.
\bibitem{Sos92}Sosnick TR, Trewhella J. Biochemistry 1992;31:8329--8335.
\bibitem{Hag98}Hagihara Y, Hoshino M, Hamada D, Kataoka M, Goto Y.
              Fold Des 1998;3:195--201.
\bibitem{Mil02}Millet IS, Townsley LE, Chiti F, Doniach S, Plaxco KW.
                Biochemistry 2002;41:321--325.
\bibitem{Cho02}Choy WY, Mulder FAA, Crowhurst KA, Muhandiram DR, 
                   Millett IS, Doniach S, Forman-Kay JD, Kay LE.
                   J Mol Biol 2002;316:101--112.
\bibitem{Shi02}Shimizu S, Chan HS. Proteins 2002;49:560--566.
\bibitem{Gol03}Goldenberg DP. J Mol Biol 2003;326:1615--1633.
\bibitem{Jew03}Jewett AI, Pande VS, Plaxco KW. J Mol Biol 2003;326:247--253.
\bibitem{Cha03}Chan HS, Shimizu S, Kaya H. Methods Enzymol 2003: in press.
\bibitem{Pac03}Vendruscolo M, Paci E. Curr Opin Struct Biol 2003;13:82--87.
\bibitem{HG03}Head-Gordon T, Brown S. Curr Opin Struct Biol 2003;13:160--167.
\bibitem{Uve02}Uversky VN, Fink AL. FEBS Lett 2002;515:79--83.
\bibitem{Plo01}Plotkin SS. Proteins 2001;45:337--345.
\bibitem{Cie02}Cieplak M, Hoang TX. Int J Mod Phys C 2002;13:1231--1242.
\bibitem{Cie03}Cieplak M, Hoang TX. Biophys J 2003;84:475--488.
\bibitem{Kli98}Klimov DK, Thirumalai D. Fold Des 1998;3:127--139.
\bibitem{Li_01}Li L, Shakhnovich EI. 
               Proc Natl Acad Sci USA 2001;98:13014--13018.
\bibitem{Fav02}Favrin G, Irb\"ack A, Wallin S. Proteins 2002;47:99--105.
\bibitem{Zho02}Zhou Y, Linhananta A. J Chem Phys 2002;117:8983--8995.

\end{thebibliography}
\end{document}